\RequirePackage{fix-cm}
\documentclass{svjour3}  % onecolumn (standard format)
\smartqed

\usepackage{graphicx}
 \usepackage{booktabs}
 \usepackage{rotating}
 \usepackage{longtable,pdflscape}
\usepackage{chngpage}
\usepackage{array}
\usepackage{graphicx,graphics,subfigure,multicol} 
\usepackage[colorinlistoftodos]{todonotes}
\usepackage[12pt]{extsizes}
\usepackage{geometry}
\usepackage{url}
\usepackage{todonotes}
\usepackage{hyperref}

\usepackage{lipsum}

\begin{document}

% \title{Preparing for Urban Diseases:~ An Open-Data Analysis of Heterogeneities in \textcolor{black}{lung cancer} Premature Mortality and Environmental/Non-Environmental Factors over 140 Toronto Neighborhoods}
% \title{Preparing for Urban Diseases: An Open-Data Analysis of Heterogeneities in \textcolor{black}{lung cancer} Premature Mortality and Associated Factors among Toronto Neighborhoods}
% \title{Preparing for Urban Diseases: An Open-Data Analysis of Heterogeneities in \textcolor{black}{Lung Cancer} Premature Mortality Rate (LCPMR) and Associated Factors among Toronto Neighborhoods}
\title{An Open-Data Analysis of Heterogeneities in Lung Cancer Premature Mortality Rate and Associated Factors among Toronto Neighborhoods}
%  \subtitle{Do you have a subtitle?\\ If so, write it here}

\titlerunning{Heterogeneities in LCPMR and Associated Factors}  % if too long for running head

\author{Zhanwei Du   \and
Jiming Liu (corresponding) \and Songwei Shan
} 

\institute{Department of Computer Science \\ Hong Kong Baptist University\\
}

% \newcommand{\zwfont}[1]{{\bf $\langle$ \begin{localsize}{12} #1 \end{localsize}$\rangle$}}
\newcommand{\zwfont}[1]{ \footnotesize{#1} }

\maketitle

\hrule
\vspace{0.5cm}

\begin{abstract}
\begin{quote}
\textit{In public health, various data are rigorously collected and published with open access. These data reflect the environmental and non-environmental characteristics of heterogeneous neighborhoods in cities.
In the present study, we aimed to study the relations between these data and disease risks in heterogeneous neighborhoods. 
A flexible framework was developed to determine the key factors correlated with diseases and find the most relevant combination of factors to explain observations of diseases through nonlinear analyses. 
Taking Lung Cancer Premature Mortality Rate (LCPMR)  in Toronto as an example, two environmental factors (green space, and industrial pollution) and two non-environmental factors (immigrants, and mental health visits) were identified in the relational analysis of all of the target neighborhoods. To determine the influence of the heterogeneity of the neighborhoods, they were clustered into three different classes. In the most severe class, two additional factors related to dwellings were determined to be involved, which increased the observation's deviance from 48.1\% to 80\%. 
The factors determined in this study may assist governments in improving public health policies. }

\end{quote}

\vspace{0.5cm}
\hrule

\end{abstract}
% \vspace{0.1cm}

\section*{{\sffamily \MakeUppercase{Introduction}}}
%Neighborhood-Level 
% \todo{Background Information, importance}
% 
% {Many diseases in neighborhoods of cities:} 
Public health researchers have shown that many diseases are correlated with geographic variations among the neighborhoods of cities. Moreover, the correlations between health and neighborhood/area characteristics and independent of population-level attributes \cite{diez2001investigating,macintyre2002place}. 
For instance, in {Boston}, the \textcolor{black}{incidence} of diabetes is correlated with ethnic disparities between different neighborhoods \cite{piccolo2015role}. %In Seoul, angiocardiopathy also shows the spatial correlation with $PM_{10}$ \cite{lim2014spatial}; 
In {Phoenix}, air pollution contributes to disparities in the \textcolor{black}{incidence} of asthma among children of different areas \cite{grineski2007incorporating}. In {New York}, environmental factors contribute to the spatial disparities of asthma \textcolor{black}{incidence} \cite{corburn2006urban}. 
% In {Toronto}, spatial disparities in {{immigration}} have been studied to explain discrepancies in the \textcolor{black}{LCPMR} in women \cite{brown2016detailed}.
\textcolor{black}{\textcolor{black}{Lung cancer} mortality was reported having a positive association with ambient nitrogen dioxide \cite{villeneuve2013cohort}. }

%  {Importance to study:} 
Therefore, it is important to explore the factors that influence the occurrence of different diseases in the neighborhoods of cities. 
%  {Why to study :} 
The determination of these factors would provide valuable insights into disease risk factors, and guide governments' actions in reducing harmful materials in human environments. For example, the Canadian government is currently in the process of banning asbestos because of asbestos-related diseases, such as \textcolor{black}{lung cancer} \cite{cbc_ban2016}.
% 

% {Rich Open data: }
The determination of disease risk factors requires numerous kinds of public-health-related data for many cities. Some of these data are rigorously collected and published with open access by government departments. For example, in New York, the Open Data Plan makes public data (such as business, education, and dwelling data) generated by various New York City agencies and organizations available for public use \cite{NYCopenData2016}. 
In Toronto, the available open data are rich and organized into 15 categories (such as culture, finance, and health) \cite{TorontoOpendata2016}. The industrial records date back to 1995 \cite{NPRI1995}. 
 {Rich open data are valuable, and their availability encourages future research to maximize their benefit to society.}
%  {Question: }
Based on these open data, we investigated how geographic factors influence diseases in the neighborhoods of cities. 

\begin{figure}
\centering
\includegraphics[width=1\textwidth]{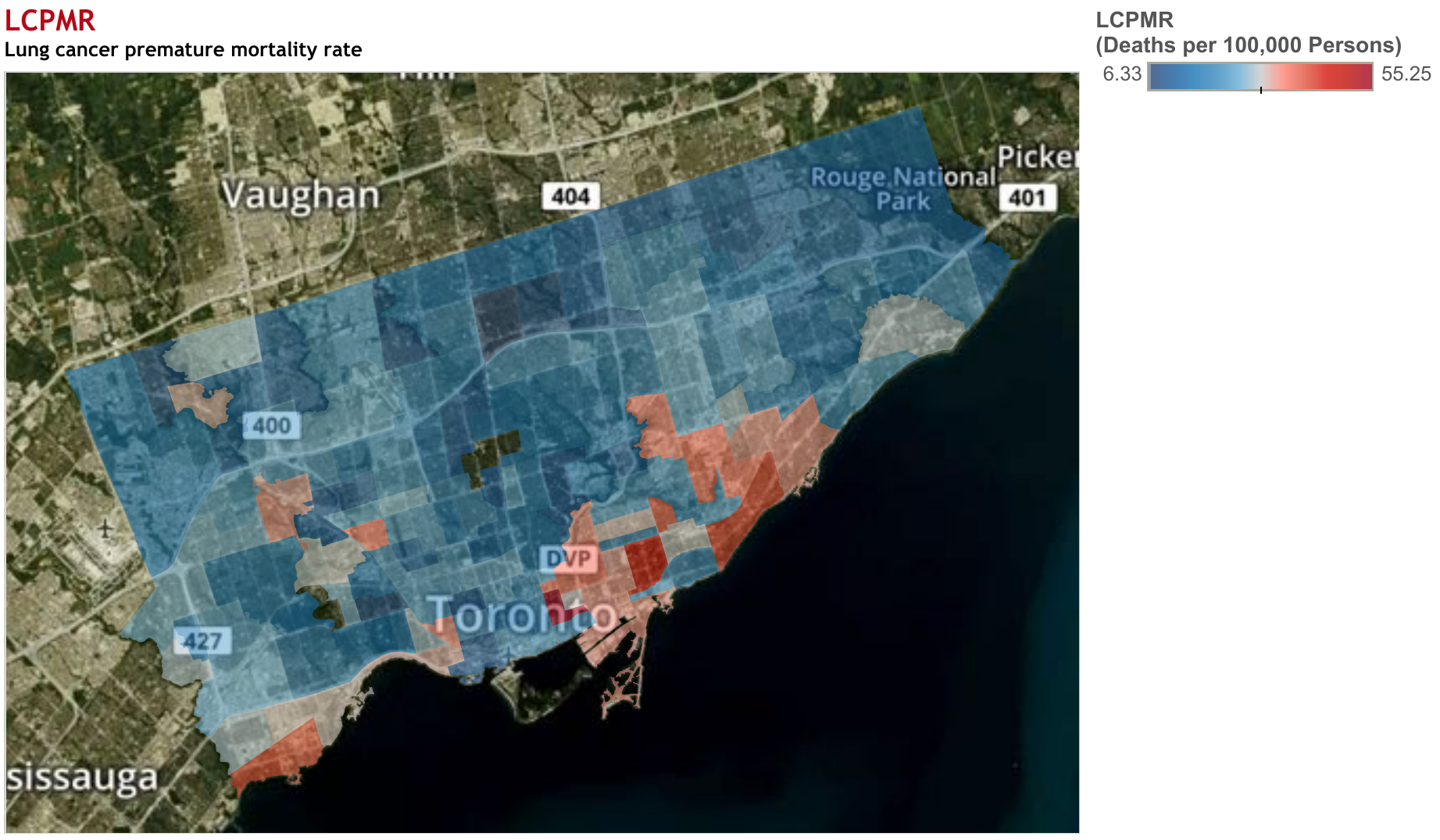}
\caption{\label{Fig:LCPMR} \zwfont{LCPMR per 100,000 persons in 138 out of the 140 neighborhoods of Toronto (2003--2007) \cite{ASD2006} (where the data from two neighborhoods were missing).
\textcolor{black}{
The spatial map was created using OpenStreetMap online platform (\href{http://http://www.openstreetmap.org/}{http://http://www.openstreetmap.org/}) (\textcolor{black}{$\copyright $ OpenStreetMap contributors}) under the license of CC BY-SA (\href{http://www.openstreetmap.org/copyright}{http://www.openstreetmap.org/copyright}). More details of the licence can be found at  \href{http://creativecommons.org/licenses/by-sa/2.0/}{http://creativecommons.org/licenses/by-sa/2.0/}.
Line graphs were drawn using Tableau Software for Desktop version 9.2.15 (\href{https://www.tableau.com/zh-cn/support/releases/9.2.15}{https://www.tableau.com/zh-cn/support/releases/9.2.15}). }
} }
\end{figure}

%  {Why choose \textcolor{black}{lung cancer} as a case study of diseases :} 
In the present study, we used \textcolor{black}{lung cancer} as a case study. Although the premature mortality rates of \textcolor{black}{lung cancer}  have decreased slowly since the late 1990s \cite{howlader2011seer}, it remains the most common cause of cancer-related death in men and the second most common in women (after breast cancer) \cite{stewart2014world,jemal2011global,ferlay2010estimates}, especially in large cities \cite{hunt2015black}. 
\textcolor{black}{
For example, worldwide, lung cancer is one of 5 most common causes of cancer death, accounting for 1.69 million deaths in 2015 \cite{WHO2017}. 
% 1.8 million people were estimated to develop lung cancer in 2013 with 1.6 million deaths \cite{stewart2014world}. 
% 
%  {Cancer's economic and death costs:}Take Canada as an example, Public Health Agency of Canada reported, in 2008, cancer was the 7th most costly illness or injury accounting for \$4.4billion in {economic costs} \cite{Burden2014canadian}. In 2015, 196,900 Canadians are estimated to {develop} cancer, and 78,000 will {die} of cancer \cite{ccssacoc2015canadian}. While in 2016, increasing populations are involved, with 202,400 Canadians and 78,800 potential deaths \cite{ccssacoc2016canadian}. 
%  {\textcolor{black}{lung cancer} is leading in cancers in {Toronto}:} 
In large cities, such as {{Toronto}},
% \textcolor{black}{lung cancer} is the second leading cause of serious disease in men and the fourth in women \cite{LC2010}. }
lung cancer is one of the most common cancers and the leading cause of cancer deaths, especially  in the low-income groups \cite{TPHLC2008,TPHLC2015}.}
The spatial distribution of the \textcolor{black}{Lung Cancer} Premature Mortality Rate (LCPMR) in the 140 neighborhoods of Toronto is depicted in Fig.\ref{Fig:LCPMR}.

%  As the most populous city of Canada, the similar situations also exist in Toronto \cite{Toronto2013}.
% 
%  {Many related factors: }
% 
% 
\textcolor{black}{Lung cancer} has been widely studied with respect to its correlation with a variety of closely related factors, such as population lifestyle \cite{hystad2014geographic}, natural environment \cite{crouse2015within}, built environment \cite{jackson2013health}, and social environment \cite{berglund2012social,casper2001definition}). 
Additionally, other {indicators} that reflect some degree of susceptibility to \textcolor{black}{lung cancer} have also been studied, such as socioeconomic status \cite{dominguez2015lung,qi2006determinants}, ethnicity \cite{jack2011lung,qi2006determinants}, and physical activity \cite{brenner2016leisure,emaus2010physical,mao2003physical}. 
{Moreover}, residents' lifestyle and work histories have been considered because of the long-term process of \textcolor{black}{lung cancer} development \cite{tomczak2016long,Lung2016,tomczak2016long}. 
% 

%  Spatial Heterogeneous
Regions and neighborhoods in urban areas generally exhibit {geographic variations} in various factors \cite{heterogeneous2011,york2011ethnic,hystad2014geographic}. For example, {{Toronto}} consists of many neighborhoods with heterogeneous environments, which exhibit variations in LCMPR \cite{TorontoCHPP2016}, lifestyle \cite{TorontoHSI2016}, demographic factors \cite{TorontoCHPP2016,TorontoCP2006,TorontoN2011}, and industry \cite{TorontoChem2012}.
% \todo[inline]{"\textcolor{black}{lung cancer} prevalence" $->$"LCMPR"}
 Hence, in the present study, we focused on the relations between \textcolor{black}{LCMPR} and heterogeneous geographic factors, based on {open data} from the neighborhoods of Toronto. 
%  To study this, 
%  Significant insights in health equality and \textcolor{black}{lung cancer} prevention for unveiling the nonlinear Influences between relevant current and past factors and \textcolor{black}{lung cancer}. 

% \todo{Prior research and the limitations, necessity}
%  {Literature Review:} 
Most studies of \textcolor{black}{lung cancer} epidemiology have focused on the relations between various factors and \textcolor{black}{lung cancer}. Broadly speaking,  {two} kinds of related factors, environmental and non-environmental, have been studied. 
% 和下文的4、6个因素关联
%  {Environmental factors}: 
Environmental factors consist of those present where a person grows, lives, works, and ages \cite{ruiz2016contributions}. Substances in the {natural, built, and social} environments may have direct interactions with the body and thereby induce kinds of diseases \cite{Protection2016}. The {natural} environment denotes naturally occurring factors, such as water, air, minerals, and climate \cite{ruiz2016contributions}. For example, \textcolor{black}{LCPMR} is related to exposure to radon and asbestos \cite{Asbestos2016,lubin1995lung}. Furthermore, air pollutants can increase the oxidative stress in cells and elevated \textcolor{black}{lung cancer} risk \cite{yang2015spatial}.
 The {built} environment refers to the manmade surroundings that provide the settings for human activity \cite{ruiz2016contributions}. For example, green space has a negative correlation with lung cancer mortality \cite{mitchell2008effect}. 
%  \todo[inline]{"\textcolor{black}{lung cancer} prevalence" $->$"\textcolor{black}{lung cancer} mortality"}
 The {social} environment encompasses social interactions and demographic factors such as dwellings and food security \cite{ruiz2016contributions}. For example, mental health has a potential correlation with \textcolor{black}{LCMPR} \cite{pirl2012depression}.

%  {Non-Environmental factors}: 
% There are two kinds of population factors (modifiable and non-modifiable) \cite{Ohio2015}. 
Non-environmental factors, such as lifestyle factors (for example, smoking and physical activity) \cite{hystad2014geographic}, ethnicity, and immigration \cite{brown2016detailed}, may correlate with disease outcomes. For example, physical activity shows a protective effect on \textcolor{black}{lung cancer risk}, with a validated biological mechanism \cite{brenner2016leisure,emaus2010physical,mao2003physical}. %For example, in {Toronto}, spatial {immigration} is also studied to explain women \textcolor{black}{lung cancer} \cite{brown2016detailed}. The influence from population non-modifiable factors is also widely studied \cite{berglund2012social,mao2001socioeconomic}. 
To determine the correlation of these factors with \textcolor{black}{lung cancer}, they are commonly modeled as linear relations using regression analyses, such as linear regression \cite{argo2010chronic} and logistic regression \cite{lopez2013lung}.
 
%  {Limitation}
%  {A few factors: } 
Research has mainly considered a small number of factors in correlation analyses. They are meaningful, but may overlook other closely related factors. Therefore, it is necessary to conduct analyses of large open datasets to determine the correlation of a large group of factors with LCMPR.
% \todo[inline]{"\textcolor{black}{lung cancer} prevalence" $->$"LCMPR"}
%  {No Heterogeneous: }
 Additionally, studies have mainly conducted city-level case studies. However, the {heterogeneity} of neighborhoods is important, as the development of neighborhoods and communities is vital to  \textcolor{black}{improve}  public health \cite{syme2009importance}.
 %{Thus it is valuable for us to identify different categories of neighborhoods with various \textcolor{black}{lung cancer} related factors, especially those neighborhoods with high \textcolor{black}{lung cancer} prevalence. }
%  {Linear Assumptions: } However, the linear assumption may be not well enough to describe the relations between them and with \textcolor{black}{lung cancer}. Thus it is worth to extend the assumption to {nonlinear} models.

%  Thus in this paper, we take Toronto as a case study, past industrial factories which have disappeared, but maybe a force not to be ignored.

% \todo{Our work}
%  {Study aim: } 
 In the present study, we investigated how various factors influence LCPMR in the neighborhoods of a city, based on {environmental and non-environmental factors from open data}, especially in neighborhoods with disproportionately high LCPMR. Specifically, we used the city of {{Toronto}} as a case study, which has a relatively high LCPMR \cite{LC2010}. Its neighborhoods are characterized by variations in LCPMR, and related factors (such as lifestyles \cite{ASD2006} and built environment \cite{TorontoN2011}).
 
%  {Study objective: }
 We analyzed the relations between various factors and LCPMR using the generalize additive model (GAM) \cite{hastie1990generalized,wood2006generalized}. GAM was used to model the nonlinearity of the relations in all of the neighborhoods.
%  {Why GAM? }
 GAM is a non-parametric regression method that is well suited to studying the relations between various factors \cite{wood2006generalized}.
To explore the influence of the heterogeneity of neighborhoods, we further explored specific neighborhoods using the same method. 

We identified neighborhoods with significant \textcolor{black}{lung cancer} factors based on open data. After dimension reduction using principal component analysis (PCA) (which has been widely used for such a purpose in various fields such as neuroscience \cite{brenner2000adaptive}), these factors were used as indicators for clustering neighborhoods. A model-based clustering method was used to cluster the neighborhoods. The data are assumed to follow a distribution that is the mixture of two or more components, which are modeled by Gaussian distributions. This method is statistically more robust than clustering algorithms based on geometric procedures (such as the k-means algorithm and hierarchical clustering algorithms) \cite{bouveyron2014model}. 
After clustering, we analyzed these factors using GAM.
%  {Why these methods: PCA ?} 
%  {Why these methods: Model-based clustering ?}

%  {Data}:
Numerous kinds of rigorously produced and published open data for Toronto are related to \textcolor{black}{lung cancer}. To sufficiently describe each neighborhood of Toronto, we collected environmental and non-environmental factors from open data. For example, 
%  {environmental factors}: 
 many environmental factors (such as soil radon potential and green space) could be found in the open demographic census \cite{TorontoCHPP2016,TorontoCP2006,TorontoN2011}. 
 In addition to commonly used factors such as green space and soil radon potential, to investigate the influence of {dwellings} on \textcolor{black}{LCPMR}, we chose the percentage of dwellings constructed before 1990, as during that period it was common to use {asbestos} for the insulation of dwellings in Toronto. 
 Moreover, to study the influence of {industry}, we considered the air pollution originating from factories that no longer existed. As \textcolor{black}{lung cancer} may remain dormant for around 20 years \cite{de2014spatial}, it is therefore common to choose a period of 20 years before the study \cite{tomczak2016long}. Hence, we used data regarding the industrial distribution in Toronto 20 years ago \cite{TorontoChem2012}), {preprocessed} by examining the effects of air pollution risks and the wind (which influences the air pollution range). 
%  {Non-environmental factors}: 
Data related to many non-environmental factors can be found in the open health surveillance indicators \cite{TorontoHSI2016}, including income, university attendance, mental health, and immigration status, to determine population lifestyles.

%  {Results: } 
%  {Regression with all the neighborhoods:} 
We first analyzed all of the neighborhoods in {{Toronto}} to determine the general relations between the target factors and LCPMR using GAM. 
%  {Results:} 
 Two \textcolor{black}{environmental}  factors (green space and industrial pollution) were derived from a combination of four factors that exhibited the highest deviance (48.1\%) among all of the neighborhoods. Along with an additional two \textcolor{black}{non-environmental} factors (immigrants and mental health visits), these factors were found to show the most significant explanatory ability of the 11 significant lung-cancer-relevant factors. 
%  {Why to Cluster?} 
 However, the deviance explained by GAM was insufficient, possibly because the heterogeneity of the neighborhoods was not considered. Therefore, we clustered neighborhoods using the model-based clustering method. 
%  {Clustering Results:} 
 The neighborhoods were clustered into three classes by \textcolor{black}{using} the model-based clustering method with the best value of Bayesian Information Criterion (BIC) as a statistical measure to \textcolor{black}{determine  unknown parameters (such as the appropriate total number of classes)}. Class A corresponded to low LCMPR neighborhoods. Classes B and C both exhibited high LCPMR,  but were featured by different levels of demographic factors. For example, the average value of median income over neighborhoods in class B was \$74,495, much higher than \textcolor{black}{that in} class C (\$44,582). 
 We chose to focus on investigating the two high-LCPMR classes, especially class C, which included enough neighborhoods for further group analysis. 
%  \todo[inline]{"\textcolor{black}{lung cancer} prevalence" $->$"LCMPR"}
%  {Regression with Class C:} 
\textcolor{black}{For classes B and C, we qualitatively and quantitatively evaluated the relations between relevant factors and LCPMR.
In view that there were only 11 neighborhoods involved in class B, we focused on the single-factor analysis to find its factor (percentage of dwellings constructed before 1990) with the highest explanatory ability (66\%) observed. 
%  {Results:}
While there were 58 neighborhoods in class C, multi-factor analysis can be applied.}
 In addition to the four most significant factors identified in the analysis of all of the neighborhoods, two dwelling factors (percentage of dwellings requiring major repairs and percentage of dwellings constructed before 1990) were also included in the group with the highest explanatory ability (80\%) observed for class C. 
%  {Comparison of All and Class C:} 
 Compared with the analysis of all of the neighborhoods, the deviance was better explained in specific neighborhoods. Specifically, more dwelling-related factors were incorporated, which increased the deviance explained from 48.1\% to 80\%.

%  {Conclusions}
 Based on public health related open data, we studied the factors influencing \textcolor{black}{LCPMR} in all of the neighborhoods. Then, to investigate the heterogeneity of the neighborhoods, we focused on factors affecting specific neighborhoods. 
%  {Method Part:} 
 A flexible framework was used to estimate the correlation of various factors with \textcolor{black}{LCPMR}, and determine which group of factors exhibited the best explanation. Clustering of neighborhoods was performed to determine their heterogeneity. 
%  {Result Part: }
 Overall, 12 factors were found to be significantly correlated with \textcolor{black}{LCPMR}. \textcolor{black}{LCPMR} was best explained by a group of four factors (immigrants, mental health visits, industrial pollution, and green space). When limiting the analysis to a class of high-risk neighborhoods, the \textcolor{black}{LCPMR} was best explained by incorporating two additional dwelling-related factors.
%  {Meaning:} 
 This research may be used to develop government guidelines for future public health policies and contribute to balancing housing, industry, and health priorities. For example, based on the {industrial and dwelling factors}, neighborhoods containing a large proportion of old houses should be prioritized, especially where industrial factories are also significant. 
 The analysis of {mental health and green space} may also support the prioritization of civic culture. 
 Although a specific disease and location were studied in this work, the methods used and questions raised in this work are broadly applicable and may motivate similar public health research.

\section*{{\sffamily \MakeUppercase{Data}}}
% \cite{TorontoLCPMR2007,Urban2013,AHD2012,ASD2006,IS2012,Smoking2007,PA2010,TorontoCP2006,TorontoN2011,TorontoChem2012,chen2008variation,NPRI1995}
To conduct our study of Toronto, we collected data regarding a variety of factors (as shown in Tab. \ref{TabVariable}) related to \textcolor{black}{LCMPR} from diverse open data sources.
Specifically, the LCMPR, the environmental factor of green space, and some non-environmental lifestyle factors (such as mental health visits, immigration status, income and university attendance) were obtained from Toronto Community Health Profiles (2006--2013) \cite{TorontoLCPMR2007,Urban2013,AHD2012,ASD2006}.
\textcolor{black}{For more details, 
LCMPR denotes the lung cancer premature mortality rate, as the premature deaths per 100,000 persons due to lung cancer \cite{TorontoLCPMR2007}. 
The factor of mental health visits denotes the percentage of people who visit the mental health service \cite{AHD2012}.
As for the green space data, it refers to the  average amount of green space per km\textsuperscript{2} in a 1-km diameter circular region \cite{Urban2013}.
And, the factors of  median income, immigrants and university attendance rate represent the percentages of immigrants, households with median  income and population with a university degree in a neighborhood, respectively \cite{ASD2006}.
There were also some non-environmental lifestyle factors (such as fruit consumption, smoking habits, and physical activities), which were obtained from Toronto’s Health Indicator Series (2007--2012) \cite{IS2012,Smoking2007,PA2010}.
 Specifically, the factors of  fruit consumption, smoking rate, and physical activity rate denote the percentages of people who consume enough daily vegetables and fruits \cite{IS2012},  smoke \cite{Smoking2007}, and have moderately active levels of physical activities \cite{PA2010}, respectively.
Other non-environmental factors (such as age) were obtained from Canada  Statistics in 2006 \cite{TorontoCP2006}. 
Specifically, the factors of three age groups from 30 to 74  refer to their populations over neighborhoods \cite{TorontoCP2006}. }

Data regarding environmental factors were related to the types of dwellings as obtained from the National Household Survey (2011) \cite{TorontoN2011}. For example,  the factors of dwellings requiring major repairs and dwellings constructed before 1990 denote the percentages of dwellings in need of major repairs and built before 1990, respectively \cite{TorontoN2011}.
  \textcolor{black}{And the natural environment factor of  soil radon potential is  to measure risk from soil radon from the study in 2008 \cite{chen2008variation}.}

We incorporated industrial influences into our study. However, most industrial facilities that were present in Toronto have recently moved elsewhere.
As \textcolor{black}{lung cancer} can remain dormant for up to 20 years \cite{de2014spatial}, we used data aggregated from industrial records from 1995 and 2012 (around 20 years before the target year of 2006). Although we lacked data from 2006, the aggregated data represented the general industrial environment of Toronto before 2006.

Industrial pollution can be spread from a pollution source by wind. Therefore, to calculate the industrial pollution, wind effects  \textcolor{black}{are} needed to be considered. The wind analysis process is detailed in the Methods section. To estimate the potential risk of industrial pollution, we obtained the toxic equivalency potentials (TEP) for relevant pollutants from the ChemTRAC Annual Report \cite{TorontoChem2012}.

\begin{figure}
\centering
\includegraphics[width=1\textwidth]{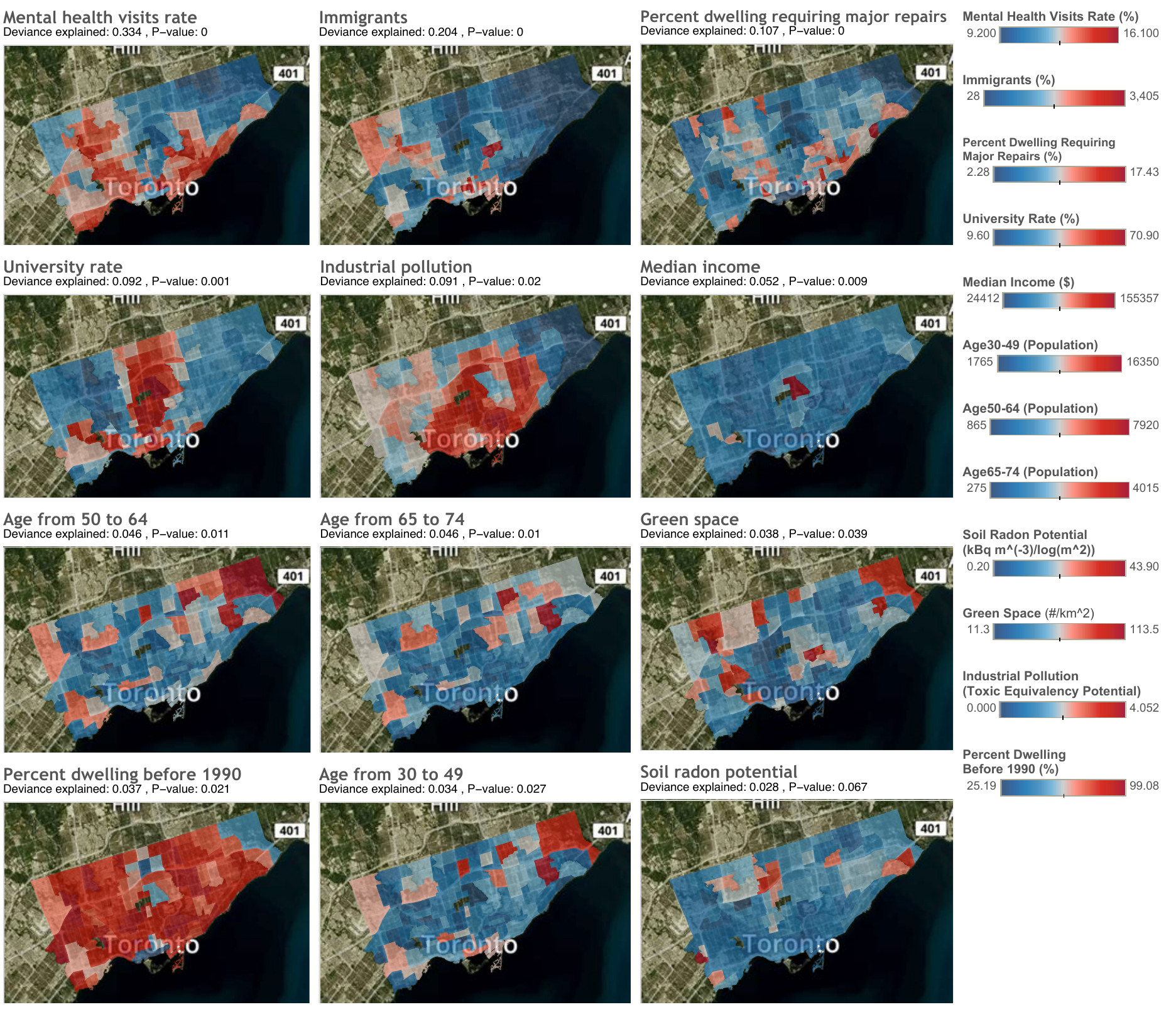}
\zwfont{ \caption{\label{Fig:factors11v2} Overview of significant factors used in the analysis of \textcolor{black}{LCPMR}. The spatial distributions determined from open data are shown. All of the factors shown are significantly correlated with LCMPR, as determined by GAM analyses. For each factor, the P-value and estimated degrees of freedom (edf) of the significance of the model smooth terms are shown. Factors are ranked by edf. Mental health visits and immigrants are the most significant factors, with deviance-explained values greater than 0.1.
\textcolor{black}{
The spatial map was created using OpenStreetMap online platform (\href{http://http://www.openstreetmap.org/}{http://http://www.openstreetmap.org/}) ($\copyright $ OpenStreetMap contributors) under the license of CC BY-SA (\href{http://www.openstreetmap.org/copyright}{http://www.openstreetmap.org/copyright}). More details of the licence can be found at  \href{http://creativecommons.org/licenses/by-sa/2.0/}{http://creativecommons.org/licenses/by-sa/2.0/}.
Line graphs were drawn using Tableau Software for Desktop version 9.2.15 (\href{https://www.tableau.com/zh-cn/support/releases/9.2.15}{https://www.tableau.com/zh-cn/support/releases/9.2.15}).}
}}
\end{figure}

Data regarding 16 factors were collected from open data sources, as mapped in Fig. \ref{Fig:factors11v2} and described in Tabs. \ref{TabVariable} and \ref{TabVariable2}, respectively. 
In the present study, 12 factors were analyzed due to significant variations in their data. The other four factors had around nine different values across the one hundred and thirty-eight neighborhoods of Toronto analyzed. Therefore, they had fewer unique covariate combinations than the specified maximum degrees of freedom of the LCMPR. Hence, we chose only 12 factors (marked by asterisks in Tab. \ref{TabVariable}) for further analysis.
As shown in Fig. \ref{Fig:factors11v2}, GAM analyses indicated that the 12 chosen factors were significantly correlated with LCMPR. The mental health visit and immigration data exhibited the most significant correlations, with deviance-explained values greater than 0.1.
% \todo[inline]{"\textcolor{black}{lung cancer} prevalence" $->$"LCMPR"}
% \begin{landscape}

% \end{landscape} [In the figure, please change 'Percent dwelling requiring major repairs' to 'Percent of dwellings requiring major repairs', 'University rate' to 'University attendance rate', and 'Percent dwelling before 1990' to 'Percent of dwellings constructed before 1990'.]

\begin{landscape}
\begin{table}[htbp]
\centering
\caption{Factors considered in the analyses of \textcolor{black}{LCPMR} in Toronto. 
 * indicates that spatial variations in the data were sufficient for further analysis. 
 {{\em All}} denotes that the variable was included in the best fitting model of all of the neighborhoods.
\textcolor{black}{ {{\em B}} and  {{\em  C}}   denote that the variable was included in the best fitting model of the neighborhoods in classes B and C, respectively.} }
%  [Please confirm whether Class A is correct, because from the text it appears that Class C was analyzed further, rather than Class A.]
\label{TabVariable}
\begin{tabular}{@{}p{4cm}p{5cm}p{2cm}p{6cm}p{1.5cm}p{2cm}@{}}
\toprule
Group       & Factor & Source(s)                                 & Description                                               & Variation sufficient & Included in best models \\ \midrule
%1
Disease  & LCPMR        & \cite{TorontoLCPMR2007}                               & \textcolor{black}{Lung cancer} premature mortality rate of premature deaths per 100,000 persons, in the period from 2003 to 2007                               & *     & {{C,All}}     \\ 
% 9
Built environment      & Industrial pollution    & \cite{TorontoChem2012,NPRI1995} &         Potential risk of industrial pollutants released by industrial facilities in either 1995 or 2012            & *     & {{C,All}}     \\
%10
Built environment     & Dwellings requiring major repairs & \cite{TorontoN2011}                              & Percentage of dwellings in need of major repairs in 2011                                  & *     & {{C}}       \\

%11
Built environment      & Green space       & \cite{Urban2013}                               & Average amount of green space (including parks and public areas) per km\textsuperscript{2} in a 1-km diameter 
% [Please confirm whether 1 km refers to the radius or diameter of this region.] 
circular region surrounding each residential block within a neighborhood in 2013.             & *     & {{C,All}}     \\

%12
Built environment      & Dwellings constructed before 1990  & \cite{TorontoN2011}                              & Percentage of dwellings built before 1990                                      & *     & {{B,C}}      \\

%13
Natural environment      & Soil radon potential    & \cite{chen2008variation}                             & Soil radon potential index to measure risk from soil radon  in 2008                                           & *     &       \\

\\ \bottomrule
\end{tabular}
\end{table}
\end{landscape}

\begin{landscape}
\begin{table*}[htbp]
\centering
\caption{Table 1 continued. Factors considered in the analyses of \textcolor{black}{LCPMR} in Toronto. * indicates that spatial variations in the data were sufficient for further analysis. 
 {{\em All}} denotes that the variable was included in the best fitting model of all of the neighborhoods.
\textcolor{black}{ {{\em B}} and  {{\em  C}}   denote that the variable was included in the best fitting model of the neighborhoods in classes B and C, respectively.} }
% \todo{Please confirm whether Class A is correct, because from the text it appears that Class A was analyzed further, rather than Class A.}
\label{TabVariable2}
\begin{tabular}{@{}p{4cm}p{5cm}p{2cm}p{6cm}p{1.5cm}p{2cm}@{}}
\toprule
Group       & Factor & Source(s)                                 & Description                                               & Variation sufficient & Included in best models \\ \midrule
%2
Non-environmental      & Mental health visits & \cite{AHD2012}                               & Percentage of persons attending mental health visits in 2007. Both sexes, age-adjusted: 20--64.                  & *     & {{C,All}}     \\
%3
Non-environmental       & Immigrants       & \cite{ASD2006}                               & Percentage of immigrants present in 2006                                       & *     & {{{{C,All}} }}     \\
%4
Non-environmental    & Age from 65 to 74     & \cite{TorontoCP2006}                              & Population of persons 65--74 years old in 2006                                      & *     &       \\
%5
Non-environmental     & Age from 50 to 64     & \cite{TorontoCP2006}                              & Population of persons 50--64 years old in 2006                                      & *     &       \\
%6
Non-environmental     & Age from 30 to 49     & \cite{TorontoCP2006}                              & Population of persons 30--49 years old in 2006                                    & *     &       \\
% 7
Non-environmental     & University attendance rate      & \cite{ASD2006}                               & Percentage of population with a University degree in 2006                                  & *     &       \\
% 8
Non-environmental       & Median income in 2006     & \cite{ASD2006}                               & Median of the weighted total net income of households in 2005, excluding institutional populations                   & *     &       \\

Non-environmental       & Fruit consumption       & \cite{IS2012}                               & Percentage of people whose daily vegetable and fruit consumption exceeded five servings in 2012                             &     &       \\
Non-environmental       & Smoking rate       & \cite{Smoking2007}                              & Percentage of people smoking regularly in 2007                                        &     &       \\
Non-environmental       & Physical activity rate    & \cite{PA2010}                               & Percentage of people having moderately active or higher levels of physical activity during their leisure and commuting time in 2010                    &     &       \\

\bottomrule

\end{tabular}
\end{table*}
\end{landscape}

\section*{{\sffamily \MakeUppercase{Methods}}}%section

\begin{figure}
\centering
\includegraphics[width=1\textwidth]{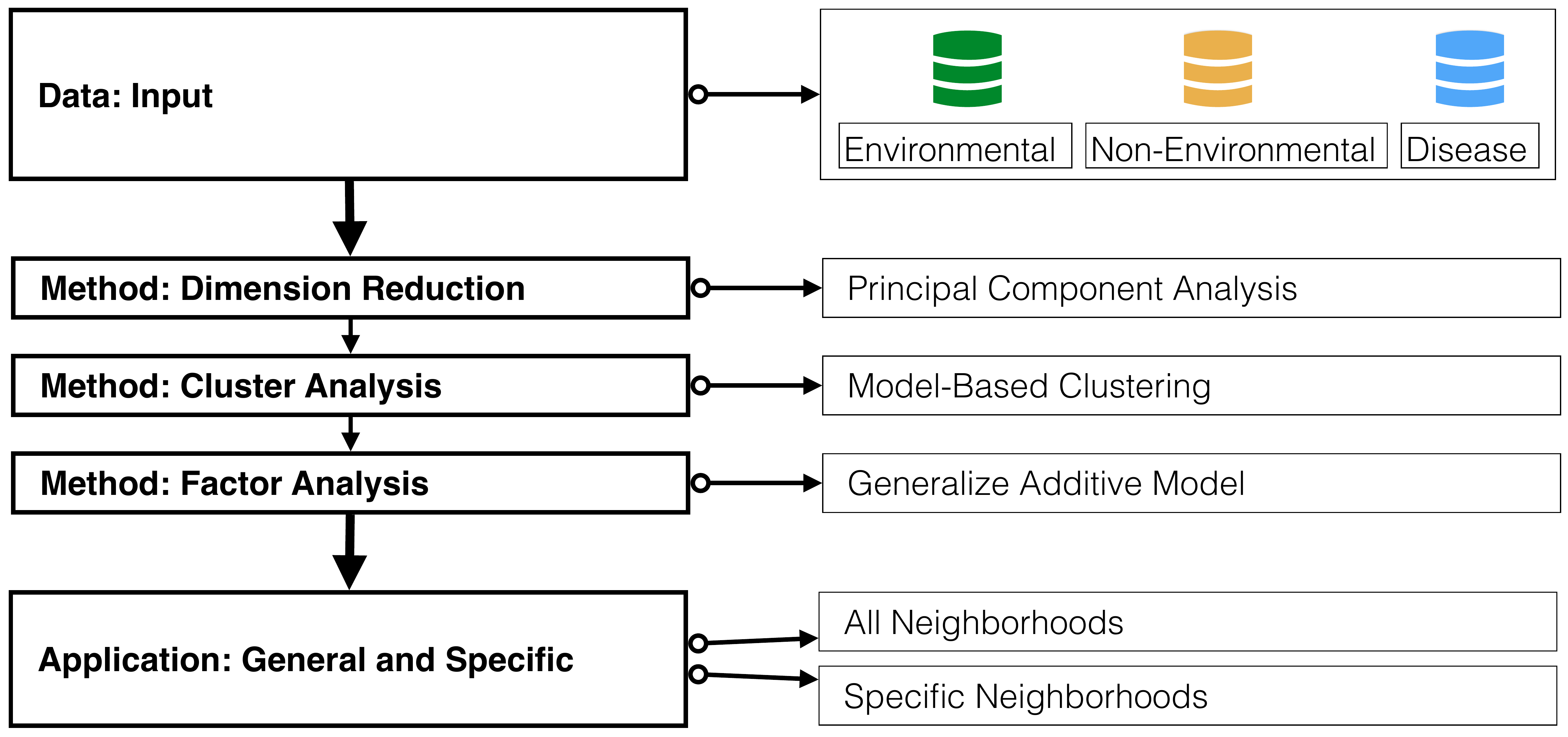}
\zwfont{\caption{\label{Fig:FigAlgorithmsProcess} Overview of the open-data analysis process. (I) Input. Data regarding the target factors were collected and preliminarily processed from available open data sources from three fields (disease, environmental, and non-environmental factors). 
(II) Methodology. The neighborhoods of the city were classified into different classes based on lung-cancer-related factors. The PCA method was used for dimensional reduction, thereby improving the results obtained from the model-based clustering method. Lastly, GAM was used to study the nonlinear relations between factors and \textcolor{black}{LCPMR}.
(III) Application. The analysis results for all or specific neighborhoods were summarized.}}
\end{figure}

To address the technical challenges of analyzing the effects of diverse factors on \textcolor{black}{LCPMR}, we used the following analysis framework. As depicted in Fig. \ref{Fig:FigAlgorithmsProcess}, in the first step, Input, data related to all of the target factors were obtained from available open data sources from three domains (environmental, non-environmental factors, and disease).  
% \todo[inline]{Change the sequence of  "disease, environmental and non-environmental factors" to "environmental, non-environmental factors and disease"}
 In the second step, Method, the neighborhoods of the city were classified based on lung-cancer-related factors. The PCA method was used for dimensional reduction, thereby improving the clustering results obtained from the model-based clustering method,  in which data's distributions were modeled by Gaussian distributions. 
 Finally, the generalize additive model (GAM) \cite{hastie1990generalized,wood2006generalized} was used to determine the nonlinear relations between factors and \textcolor{black}{LCPMR} in all or specific neighborhoods.
 In the third step, Application, the analysis results of all or specific neighborhoods were summarized.

\subsection*{{\sffamily \textbf{Industrial data preprocessed using a wind model}}}%subsection
In the Toronto case study, it was necessary to preprocess the industrial data using a wind model. 
Current industrial data indicate the health risks of air pollutants released by factories. However, the spatial range of air pollutants is affected by the wind. Therefore, it is necessary to map the risks of industrial air pollutants onto neighboring spatial locations.
We considered two factors in the spreading of pollutants: (1) the potential risk of the pollutants released and (2) the spatial range of the air pollutants. 
Specifically, a Gaussian model was used to estimate air pollutant dispersion \cite{Outline2016}. The pollutant distribution was assumed to follow a Gaussian probability distribution \cite{ADMS2016}.
The spatial range of pollutants was influenced by wind direction. The potential risk at location $B$ from factory $A$, $R_{F_A \rightarrow L_B}$, could be modelled as follows:
% $Risk_{A2B} \sim log(Riskcancer*Pdensity*Pwind_direction)$
\begin{equation}
R_{F_A \rightarrow L_B} \sim log \{ R_{TEP}^{A} * P_{Density}^{A \rightarrow B}*P_{Wind}^{A \rightarrow B} \} 
\end{equation}
where $TEP$ refers to the Toxic Equivalency Potential (TEP) as the potential risks of the pollutants to cause harm \cite{TorontoChem2012}. The total $TEP$ from various pollutants released by factory $A$,  $R_{TEP}^A$, was defined as follows:
\begin{equation}
 R_{TEP}^A = \sum_{i} V_{TEP_i}^A
 \end{equation}
where $V_{TEP_i}^A$ is the potential risk of the $i$-th pollutant released by factory $A$. The density of pollutants $P_{Density}^{A \rightarrow B} $ released by factory $A$ to location $B$ was assumed to follow a Gaussian probability distribution:
\begin{equation}
P_{Density}^{A \rightarrow B} = Gaussian(\mu, \sigma)
 \end{equation}
where $\mu$ is the range expectation of highest density, usually within the closest 5 km \cite{argo2010chronic}. The variance $\sigma$ was set as 1, as the integer which produced the best correlation between the spatial distribution of air pollutants and \textcolor{black}{LCPMR}. % as high as 64.5\%.

\subsection*{{\sffamily \textbf{Classification methods}}}%subsection
To analyze the \textcolor{black}{LCPMR} in specific neighborhoods, we classified the neighborhoods based on the target factors to determine their heterogeneity. The classification process consisted of two steps: (1) Dimensional reduction was conducted using PCA on the feature vector, and represented by the first two principals. The percentages of the total variance explained by the two principals were as high as around 98\%.
 (2) Using the two principals of the feature vector, the model-based classification method was used for classification of neighborhoods in Toronto  \textcolor{black}{ by inferring class categories through the estimation of Bayesian Information Criterion (BIC) (a statistical measure to compare  models of different parameter sets)}. 

\subsection*{{\sffamily \textbf{Nonlinear relation analysis method}}}%subsection
To evaluate how various factors were correlated with \textcolor{black}{LCPMR}, GAM \cite{hastie1990generalized,wood2006generalized} was used to model their nonlinear relations in all and specific neighborhoods. 
Specifically, $Lc_i$, the \textcolor{black}{LCPMR} in the $i$-th location was related with specific factors through a link function $g$ as follows:
 \begin{equation}
g(Lc_{i})= \sum_{k=1}^{K} f_k(x_k^i)
 \end{equation}
where $x_k^i$ represents the $k$-th feature of the $i$-th location, and the function $f_k$ is a smooth function.

\subsection*{{\sffamily \textbf{Factor analysis}}}%subsection
We analyzed the synthetic nonlinear relations of various factors with \textcolor{black}{LCPMR} in Toronto and in specific neighborhoods. 
However, similar factors may exhibit strong linear relations with each other. Therefore, to reduce interference, we chose to use one of the three significant linearly correlated age factors (percentages of the population aged from 30 to 49, 50 to 64, or 65 to 74), in all of the analyses, and one of the two dwelling-related factors (either percentage of dwellings constructed before 1990 or percentage of dwellings needing major repairs) in the analysis of all of the neighborhoods.
Factors with P-values \textcolor{black}{larger} than 10\% in the synthetic analysis were discarded. The final set of factors consisted of the group with the highest explanation capacity.

\section*{{\sffamily \MakeUppercase{Results}}}%section
In this section, we describe the results of the analyses of all of the neighborhoods and specific classes of neighborhoods. % To conduct these analyses, we used the GAM package \cite{trevorhastie2016} for R software.

\subsection*{{\sffamily \textbf{Factor analysis of all neighborhoods}}}%subsection
We first analyzed all of the neighborhoods in {{Toronto}} to determine the general relations between various factors and \textcolor{black}{LCPMR} using GAM. 
% To characterize factors' influence on \textcolor{black}{lung cancer} prevalence, we start by analyzing all neighborhoods with nonlinear relations among all factors and \textcolor{black}{lung cancer} prevalence rate.

\begin{figure}[ht!]
\centering
\includegraphics[width=1\textwidth]{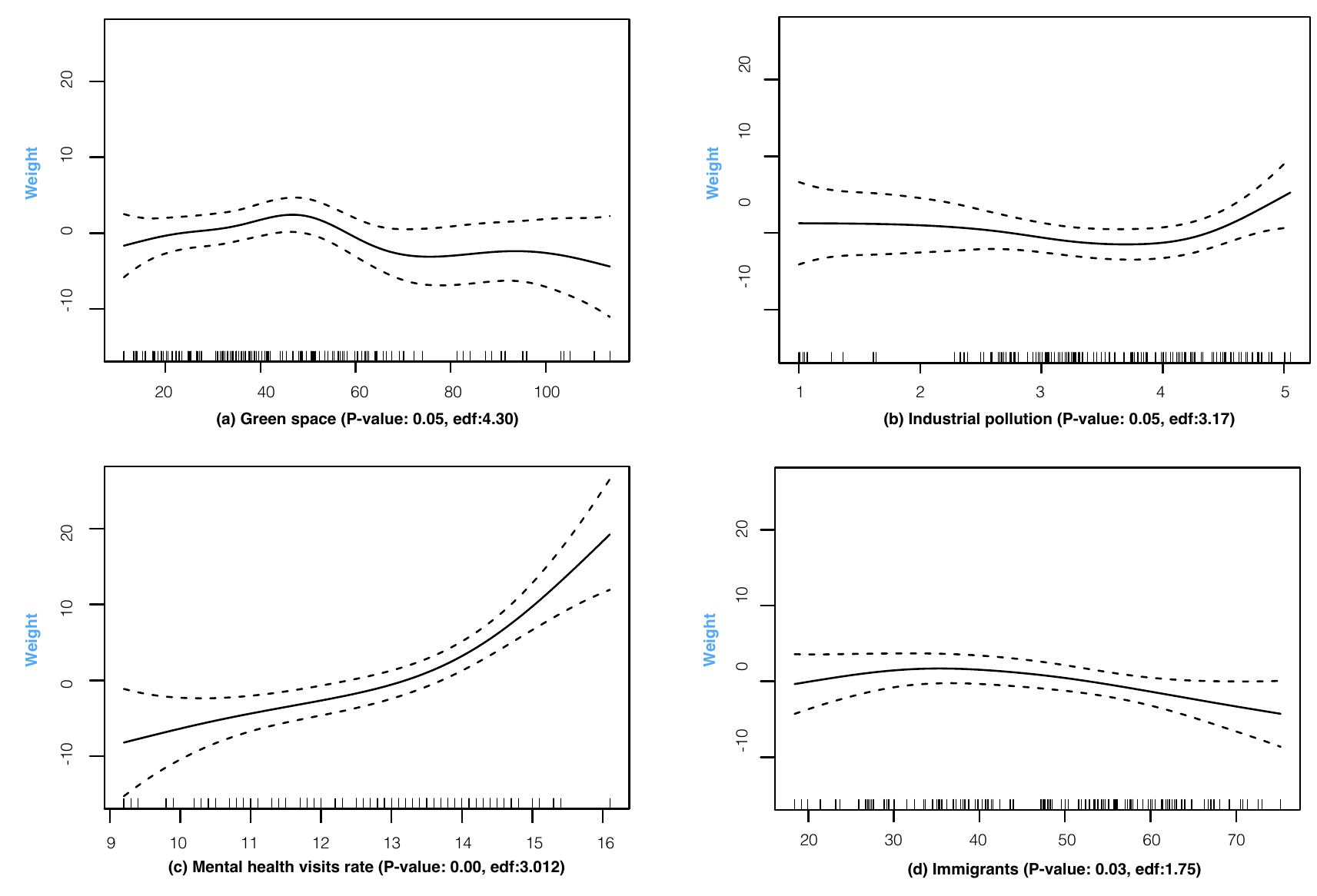}
\zwfont{ \caption{\label{Fig:Figcity_levelCity} Overview of nonlinear relations between various factors and LCMPR in all of the neighborhoods. The mean values of the model smooth term are plotted with solid lines, and the confidence intervals are plotted with dashed lines. In addition, the P-value and edf of the significance of the model smooth terms are shown on the x-axes. The overall deviance explained is 0.481.}}
\end{figure}

The nonlinear analysis results are shown in Fig. \ref{Fig:Figcity_levelCity}. The group that best explains the observed LCMPR in Toronto comprise four factors. Two of these factors (industrial pollution and mental health visits) correlate positively with \textcolor{black}{LCPMR}. Conversely, the other two factors (immigrants and green spaces) are negatively correlated. 
% \todo[inline]{"\textcolor{black}{lung cancer} prevalence" $->$"LCMPR"}
Initially, the nonlinear influence of {industrial pollution} is stable, then increases after reaching a threshold value of around 4. This suggests it is necessary to maintain industrial pollution levels below this threshold.
Analysis of {green space} indicated that its influence on \textcolor{black}{LCPMR} increased significantly after exceeding a threshold of around 50.

These results suggest that general trends in the \textcolor{black}{LCPMR} can be explained by the following four factors: mental health, immigration, industrial pollution, and green space.

\subsection*{{\sffamily \textbf{Factor analysis of specific neighborhoods}}}%subsection
\begin{figure}
\centering
\includegraphics[width=1\textwidth]{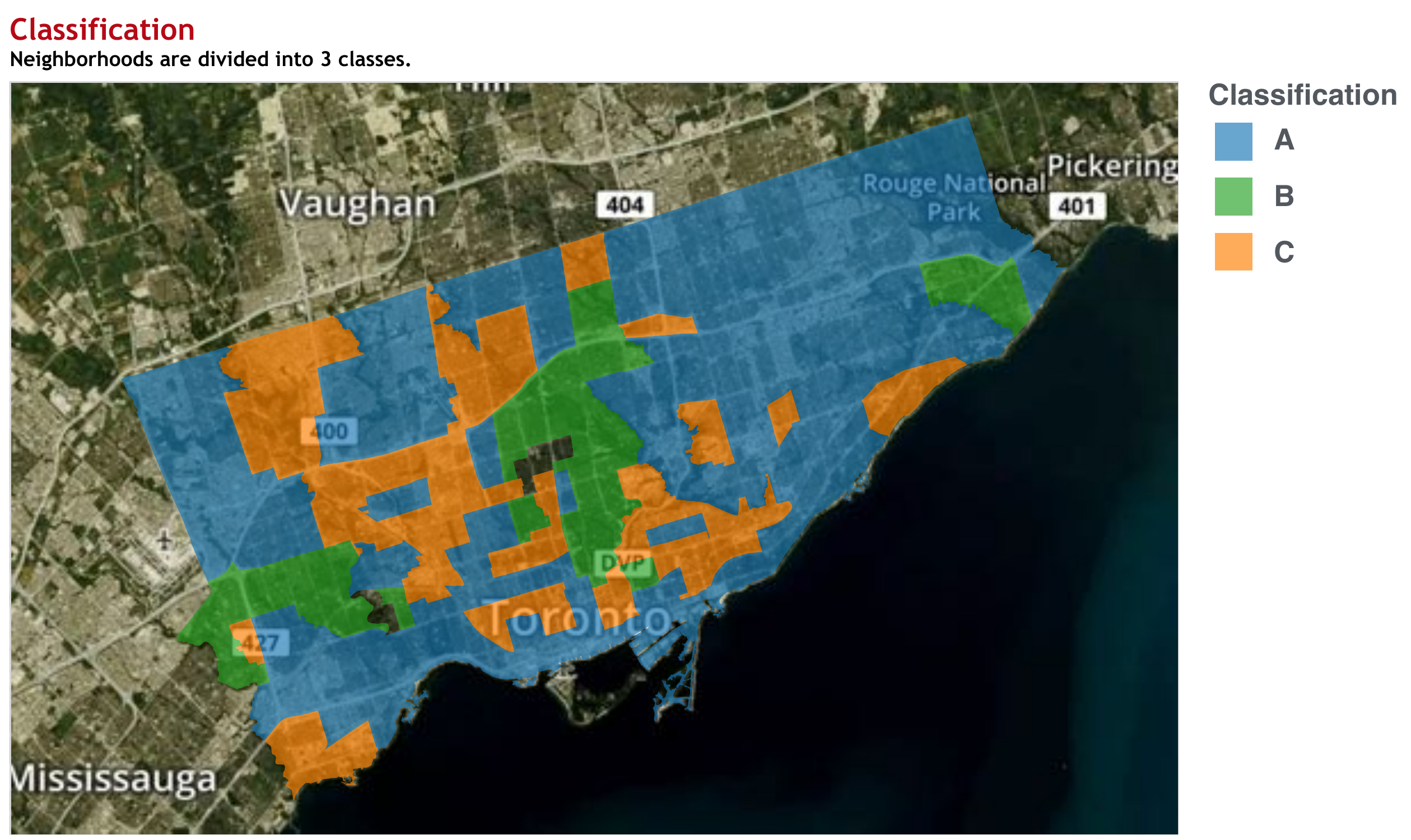}
\zwfont{\caption{\label{Fig:FigClassification} Overview of classification results. The three classes were identified using an classification method to find the best \textcolor{black}{ Bayesian Information Criterion (BIC) (a statistical measure to compare  models of different parameter sets)}.  There were 61 neighborhoods in class A, 11 in class B and 58  in class C . 
The spatial map was created using OpenStreetMap online platform (\href{http://http://www.openstreetmap.org/}{http://http://www.openstreetmap.org/}) ($\copyright $ OpenStreetMap contributors) under the license of CC BY-SA (\href{http://www.openstreetmap.org/copyright}{http://www.openstreetmap.org/copyright}). More details of the licence can be found at  \href{http://creativecommons.org/licenses/by-sa/2.0/}{http://creativecommons.org/licenses/by-sa/2.0/}.
Line graphs were drawn using Tableau Software for Desktop version 9.2.15 (\href{https://www.tableau.com/zh-cn/support/releases/9.2.15}{https://www.tableau.com/zh-cn/support/releases/9.2.15}).
The layouts were modified with Keynote version 6.6.2 (\href{http://www.apple.com/keynote/}{http://www.apple.com/keynote/}). 
}}
\end{figure}

\begin{figure}
\centering
\includegraphics[width=0.8\textwidth]{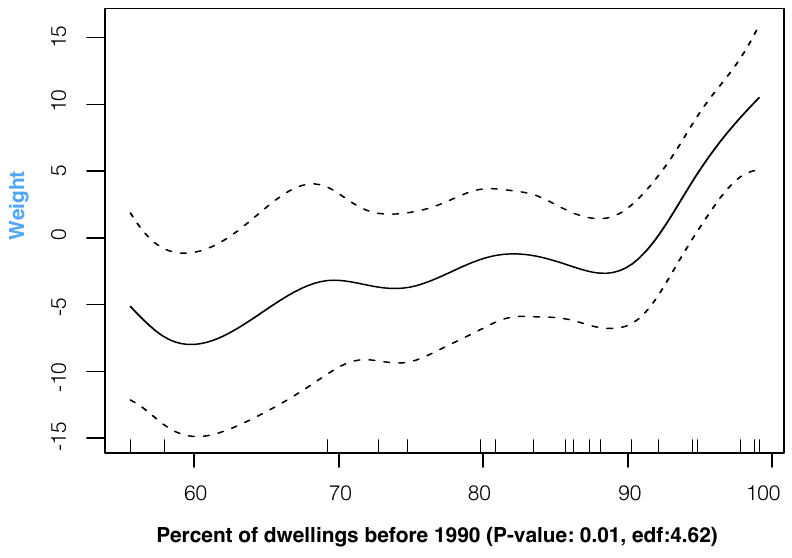}
\zwfont{\caption{\label{Fig:FigClassB} Overview of nonlinear relation between the most significant factor and LCMPR in class B.
% [Please confirm whether class A is correct, because from the text it appears that class C was analyzed further, rather than class A.] 
The mean values of the model smooth terms are plotted with solid lines, and the confidence intervals are plotted with dashed lines. The P-value and edf of the significance of the model smooth terms are shown on the x-axes. 
The total deviance explained is 0.663. }}
% [In the figure, please change 'Percent dwelling requiring major repairs' to 'Percent of dwellings requiring major repairs' and 'Percent dwelling before 1990' to 'Percent of dwellings constructed before 1990'.]
\end{figure}

\begin{landscape}
\begin{figure}
\centering
\includegraphics[width=1.4\textwidth]{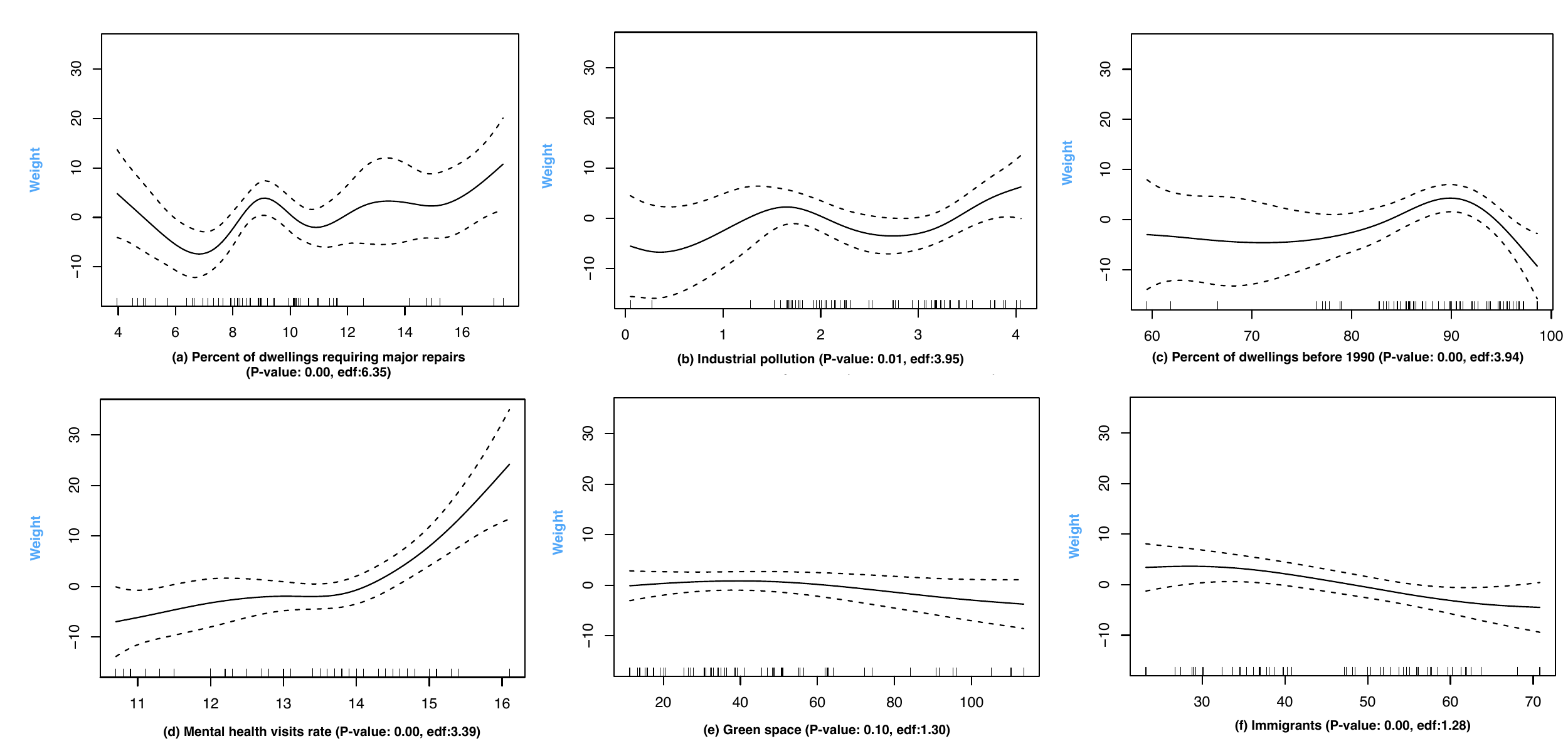}
\zwfont{\caption{\label{Fig:FigClassA} Overview of nonlinear relations between various factors and LCMPR in class C.
% [Please confirm whether class A is correct, because from the text it appears that class C was analyzed further, rather than class A.] 
The mean values of the model smooth terms are plotted with solid lines, and the confidence intervals are plotted with dashed lines. The P-value and edf of the significance of the model smooth terms are shown on the x-axes. 
The total deviance explained is 0.807. }}
% [In the figure, please change 'Percent dwelling requiring major repairs' to 'Percent of dwellings requiring major repairs' and 'Percent dwelling before 1990' to 'Percent of dwellings constructed before 1990'.]
\end{figure}
\end{landscape}

The deviance explained by the overall analysis was significant but not fine-grained, as the heterogeneity of the neighborhoods was not directly addressed.
To further investigate this heterogeneity, the neighborhoods were grouped into three classes identified using an automatic classification method (detailed in the Methods section). 
As shown in Fig. \ref{Fig:LCPMR}, neighborhoods in Toronto are characterized by different levels of \textcolor{black}{LCPMR}. Therefore, we classified these neighborhoods using a model-based classification method to identify different classes of neighborhoods automatically. The classification results are shown in Fig. \ref{Fig:FigClassification}. 
Three classes were identified with \textcolor{black}{the best  BIC}. Class A was characterized by low levels of \textcolor{black}{LCPMR}; the other two classes (B and C) exhibited high levels of \textcolor{black}{LCPMR}, \textcolor{black}{ but with different levels of demographic factors.
 For example, the median income over neighborhoods in class B was \$74,495 on average, much higher than that in class C (\$44,582)}. 
Since class B contained only a small number of neighborhoods,   \textcolor{black}{we focused on the single-factor analysis to find its factor with the highest explanatory ability observed.  As depicted in Fig. \ref{Fig:FigClassB}, the factor  of percentage of dwellings constructed before 1990 was found to be positively correlated with LCPMR, which gave the deviance-explained value as high as 66\%.  
In addition, we chose class C for multi-factor analysis.}
The analytical results for class C are shown in Fig. \ref{Fig:FigClassA}. The four most significant factors determined in the overall analysis were also significant in class C. Also, two dwelling-based factors were present in the group that best explained the observed \textcolor{black}{LCPMR}. 
% \todo[inline]{"\textcolor{black}{lung cancer} prevalence" $->$"LCMPR"}
Specifically, the percentage of dwellings needing major repairs were positively correlated with \textcolor{black}{LCPMR}. The significance of the percentage of dwellings constructed before 1990 initially increased with the percentage, then decreased after reaching a threshold value.

These findings suggest that the heterogeneity of neighborhoods should not be neglected. The significance of different factors may vary between neighborhoods, which may not be evident in general analysis. Moreover, the deviance is explained better in analyses of specific neighborhoods than in general analyses. In our Toronto case study, the deviance-explained values increased from 48.1\% to 80\%  \textcolor{black}{when the heterogeneity was considered in the multi-factor analysis.}

\section*{{\sffamily \MakeUppercase{Discussion}}}%section

% To study the impacts of various factors on the \textcolor{black}{lung cancer} prevalence is important in public health to 
% reduce the serious threat of \textcolor{black}{lung cancer}, especially from the huge amount open data.
% Taking Toronto as a case study, by investigating the impacts of various factors on \textcolor{black}{lung cancer} from open data, we have shown their nonlinear relations in all and specific neighborhood levels. 
We investigated how various factors, analyzed using open data, were connected with disease risks in different neighborhoods of a city.
Using Toronto as a case study, we investigated the effects of various factors on \textcolor{black}{Lung Cancer Premature Mortality Rate (LCPMR)}, and their nonlinear relations were analyzed across the whole city and in specific neighborhoods. 

 {Green space} has been shown to promote physical activity and mental health \cite{humpel2002environmental,cohen2015access}. Hence, the presence of green space may improve individuals' health and reduce the LCPMR.
 {Industrial pollution} released by factories shows a significant correlation with \textcolor{black}{LCPMR}, which may be attributed to the resultant air pollution \cite{argo2010chronic}. For example, in Canada, people residing in areas exposed to airborne pollutants from industrial releases in {1967{-}1970} have a higher risk of cancers.

 Of the 11 significant lung-cancer-relevant factors identified in the present study, two factors, immigrants and mental health visits, were the most significant factors in determining the LCPMR. 

 {Mental} health may be correlated with the LCPMR \cite{pirl2012depression}, which may result from the influence of depression on people's health behavior and their use of health care \cite{shimizu2012clinical}.
%  \todo[inline]{"\textcolor{black}{incidence} of \textcolor{black}{lung cancer}" $->$"\textcolor{black}{lung cancer} Premature Mortality Rate "}
 The lower \textcolor{black}{LCPMR} in areas with higher immigrant populations may be attributed to  \textcolor{black}{ our studied period, which cannot reflect the long-term trend of immigrants' lung cancer fully.   People needed  up to 20 years to develop lung cancer \cite{de2014spatial}. While the immigrants that were studied in this work only settled in Toronto for more than one year in 2006, as compared with the monitored LCPMR  period  from 2003 to 2007.  }%Thus, we observed the increasing trend of LCPMR along with increasing immigrants, which have little chance to develop lung cancer.}
%  As lung cancer can remain dormant for up to 20 years \cite{de2014spatial}, we used data aggregated from industrial records from 1995 and 2012 (around 20 years before the target year of 2006). 
%  {immigrants}' health being generally better than the health of Canadian-born citizens \cite{ng2013dynamics,ng2011healthy,pottie2008language,newbold2005self}. 

 Due to the heterogeneity among different neighborhoods, the general relations between factors and \textcolor{black}{LCPMR} may not well explain the \textcolor{black}{LCPMR} in specific neighborhoods. Thus, it is valuable to determine relations within specific neighborhoods. For example, in class C, the group of factors that best explained the \textcolor{black}{LCPMR} contained more factors (six, rather than four) and the deviance-explained value of the best group increased from 48.1\% to 80\%, compared with the general analysis.

 The two additional factors in the class C analysis were both {dwelling related}, perhaps due to the increased exposure to radon and asbestos in these neighborhoods' dwellings \cite{Asbestos2016,lubin1995lung}. For instance, dwellings that require major repairs are more likely to have higher concentrations of radon due to inadequate ventilation \cite{RadonBuy2016}. Furthermore, dwellings constructed in Toronto before 1990 may contain {asbestos} in insulation fabricated from vermiculite \cite{AsbestosFAQs2016}.

As indicated by comparing the overall analysis and those of specific neighborhoods, geographic heterogeneity should be considered in further research and in public health policy.

The results obtained indicate the correlational relations between various factors and \textcolor{black}{LCPMR}. However, they do not indicate the corresponding causal relations. 
Hence, causal relations should be examined in future studies. 
Additionally, when calculating the influence of the wind on industrial pollution, the parameters in the wind model were assumed to be homogeneous for all of the industrial facilities across different neighborhoods. 
Although this limitation may result in some degree of inaccuracy, the measurements reflect the general disparities between Toronto neighborhoods, and therefore the correlations are useful and significant.

The present study provides a flexible framework to investigate the influence of various factors on diseases in cities and in particular neighborhoods thereof. Although the disease and methods used may be particular to the Toronto case study, the questions raised in this study are valuable for similar research in public health using open data. 
Moreover, this research may be useful to governments in developing future public health policies, and to balance housing, industrial, and health priorities.

\section*{{\sffamily \MakeUppercase{Conclusion}}}%section
Based on diverse factors derived from rich open data, the present study aimed to determine their relations to \textcolor{black}{ Lung Cancer Premature Mortality Rate (LCPMR)} in the heterogeneous neighborhoods of Toronto.  
To achieve these aims, a flexible framework was developed to recognize potential factors nonlinearly correlated with disease and to determine a set of factors that best explained the observed disease \textcolor{black}{LCPMRs}. In addition, clustering of neighborhoods was used to determine heterogeneity. 

Taking Toronto as a case study, the convergence of the main factors was used to explain the observed LCPMRs in heterogeneous neighborhoods. Specifically, four significant factors (green space, industrial pollution, immigrants, and mental health visits) were identified in the relational analysis of all of the neighborhoods. To determine the heterogeneity of Toronto's neighborhoods, we clustered the neighborhoods into three different classes, and conducted an in-depth analysis of the class with the highest LCMPR. Two additional dwelling-related factors were incorporated, which increased the deviance explained from 48.1\% to 80\%.

The factors influencing diseases in cities become increasingly sophisticated with the development of new physical and social environments. It is possible that some emerging factors that contribute to diseases, such as \textcolor{black}{lung cancer}, have not yet been identified. 
Alternatively, some factors traditionally thought to be unrelated to disease may increase in significance as a result of changes in complex environmental conditions. 
Based on advances in open data, interdisciplinary collaborations in the analyses of various factors may determine their influences on diseases in heterogeneous regions. The results of this research may guide governments in developing public health policies.

\bibliographystyle{ama}
\bibliography{template}

\end{document}